\documentstyle[preprint,revtex,eqsecnum]{aps}

\begin{document}
\preprint{IFT-P.008/93}
\preprint{January 1993}
\preprint{hep-ph/9302201}

\begin{title}
Neutrinoless double beta decay in an \\ $SU(3)_L\otimes U(1)_N$
model
\end{title}
\author{ V. Pleitez  and M.D. Tonasse}
\begin{instit}
Instituto de F\'\i sica Te\'orica\\
Universidade Estadual Paulista\\
Rua Pamplona, 145\\
01405-900-- S\~ao Paulo, SP\\
Brazil
\end{instit}
\begin{abstract}
We consider a model for the electroweak interactions with
$SU(3)_L\otimes U(1)_N$ gauge symmetry. We show that, it is the
conservation of $F=L+B$ which forbids massive neutrinos and the
neutrinoless double beta decay, $(\beta\beta)_{0\nu}$. Explicit and
spontaneous breaking of $F$ imply that the neutrinos have an
arbitrary mass and $(\beta\beta)_{0\nu}$ proceeds also with some
contributions that do not depend explicitly on the neutrino mass.
\end{abstract}
\pacs{PACS numbers: 12.15.Cc; 12.15.Ji}
\narrowtext
\section{Introduction}
\label{sec:intro}
It is well known that the observation of neutrinoless double
beta decay, will imply a new physics beyond
the standard model. Usually, two kinds of mechanisms for
this decay were independently assumed: massive Majorana neutrinos
and right-handed currents~\cite{d}.

In the latter case, neutrino mass is not required but, if the
right-handed currents are part of a gauge theory, it has been
argued that at least some neutrinos must have non-zero masses. If
this is the case, the observation of $(\beta\beta)_{0\nu}$
would imply the existence of massive neutrinos, whether or not
right-handed currents exist~\cite{bk}. It is also well known that
whatever mechanism generates the neutrinoless
double beta decay, it also generates a Majorana
mass term~\cite{sv,t}. In this sense, the fundamental concept underlying
that decay is the existence of massive neutrinos. However an
important point is to find out what is the mechanism which triggers
$(\beta\beta)_{0\nu}$ and this could not be the neutrino mass.

Models based on the gauge symmetry $SU(3)_c\otimes SU(3)_L\otimes
U(1)_N$ are interesting extensions of
the Standard Model. It is possible, in this class of models
to cancel anomalies if the number of families are divisible by the
number of colors~\cite{v,vs,pp}. The new gauge sector
of the model must be very massive because there are flavor changing
neutral currents mediated by the new neutral gauge boson, $Z'^0$.
This implies that $M_{Z'}>40$ TeV~\cite{pp}.

Some time ago it was claimed that in this sort of models
$(\beta\beta)_{0\nu}$ could occur even with massless
neutrinos~\cite{pp1}. Here we return to this
question, and show that the neutrinos could remain massless but not
in a {\em natural} way.

Here, the word {\em natural} is used in the technical
sense~\cite{gg}. It means that the masses in a theory
are {\em finite} and {\em calculable} if there is a
zeroth-order mass relation which is invariant under
arbitrary changes of parameters of the theory. In fact, in
renormalizable theories a mass is either zero because some unbroken
symmetry or arbitrary since it is necessary a counterterm in order
to implement the renormalization program, leaving masses as free
parameters of the theory. Hence, if a mass is zero at tree level, and
there is no a respective counterterm, loop corrections cannot
be divergent since there is no counterterm available to cancel the
infinities.

In fact, there are contributions to
the no-neutrino $\beta\beta$ decay that do not depend explicitly on
the neutrino mass, but it is not possible
to keep the neutrino massless, at least not in the {\em natural} way
we have discussed above.
The present necessity of massive neutrinos has no relation with the
bad high energy behavior of processes as $W^-V^-\to
e^-e^-$~\cite{bk} since in this models the doubly charged gauge boson
$U^{--}$ cancels the divergent part of such a process.

If one wants lepton number to be conserved, one must assume that the
$V^\pm,U^{\pm\pm}$ gauge bosons carry lepton number. The lepton number
conservation can also still be maintained in the Yukawa sector by
assigning appropriate lepton number to the scalar fields too.
This naivy assigment of lepton number does not work because
the vector bosons also couple to the quarks. A more appropriate
quantum number is the {\em lepto-baryon}
number, $F=L+B$ defined in Sec.~\ref{sec:model}. It is the
conservation of this quantum number which forbids massive neutrinos and
$(\beta\beta)_{0\nu}$. We add trilinear terms to the Higgs potential,
Sec.~\ref{sec:scalar}, which explicitly violate $F$ and in
Sec.~\ref{sec:decay} we consider the $(\beta\beta_{0\nu})$
decay. In Sec.~\ref{sec:con} we show that the explicit $F$-violation
implies also an spontaneously breaking of this quantum number and
neutrinos with arbitrary mass.

\section{The $SU(3)_L\otimes U(1)_N$ Model}
\label{sec:model}
Let us first recall some points of the model of Ref.~\cite{pp}.
The representation content is the following: the leptons transform as
triplets,
\begin{equation}
\psi_{lL}=\left(\begin{array}{c}
\nu_a\\ l_a\\ l^c_a\end{array}\right)_L\sim({\bf3},0),
\label{leptons}
\end{equation}
with $a=e,\mu,\tau$. In the quark sector we have the triplet
\begin{equation}
Q_{1L}= \left(\begin{array}{c}
u_1\\ d_1 \\ J_1\end{array}\right)_L\sim({\bf3},+\frac{2}{3}),
\label{q1}
\end{equation}
for the left-handed fields, and singlets
\begin{equation}
u_{1R}\sim({\bf1},+\frac{2}{3});\;
d_{1R}\sim({\bf1},-\frac{1}{3});\;
J_{1R}\sim({\bf1},\!+\frac{5}{3}),
\label{rh}
\end{equation}
for the respective right-handed fields, and we have not introduced
right-handed neutrinos.

The second and third families of quarks are in antitriplets
$({\bf3}^*,-\frac{1}{3})$
\begin{equation}
Q_{2L} = \left(
\begin{array}{c}
J_2\\
u_2\\
d_2
\end{array}\right)_L,
Q_{3L} = \left(\begin{array}{c}
J_3\\
u_3\\
d_3
\end{array}\right)_L.
\label{q23}
\end{equation}
The respective right-handed quarks are also in singlets. In fact,
two of the three quark generations, it does not matter which,
transform identically in contrast to the third one. The model is
anomaly free if we have equal number of triplets and antitriplets,
counting the color of $SU(3)_c$, and furthermore requiring the sum of
all fermion charges to vanish. The anomaly cancellation occurs for
the three generations together and not generation by generation. In
Eqs.~(\ref{q1}) and (\ref{q23}) the quarks, except the one with
charge $+\frac{5}{3}$, are linear combinations of the mass
eigenstates.

For the first generation of quarks we have the following charged
current interactions:
\begin{equation}
{\cal L}^{CC}_{Q_1W}=-\frac{g}{\sqrt2}\left(\bar u_L\gamma^\mu
d_{\theta L}W^+_\mu+\bar J_{1L}\gamma^\mu u_LV^+_\mu+\bar d_{\theta
L}\gamma^\mu J_{1L}U^{--}+H.c.\right),
\label{ccq1}
\end{equation}
and, for the second generation of quarks we have
\begin{equation}
{\cal L}^{CC}_{Q_2W}=-\frac{g}{\sqrt2}\left(\bar c_L\gamma^\mu
d_{\theta L}W^+_\mu-\bar s_{\theta L}\gamma^\mu J_{2\phi L}V^+_\mu+\bar
c_L\gamma^\mu  J_{2\phi L}U^{++}+H.c.\right).
\label{ccq2}
\end{equation}
The charge changing interactions for the third generation of quarks
are obtained from those of the second generation, making
$c\rightarrow t$, $s\rightarrow b$ and $J_2\rightarrow J_3$.
We have mixing only in the $Q\!\!=\!\!-\frac{1}{3}$ and
$Q\!\!=\!\!-\frac{4}{3}$ sectors, then in Eqs.~(\ref{ccq1}) and
(\ref{ccq2}) $d_\theta,s_\theta$ and $J_{2\phi}$ mean
Cabibbo-Kobayashi-Maskawa states in the three and two-dimensional
flavor space $d,s,b$ and $J_2,J_3$ respectively. In the leptonic
sector we have the charged currents
\begin{equation}
{\cal L}_l^{CC}=-\frac{g}{\sqrt2}\sum_l\left(\bar\nu_{lL}\gamma^\mu
l_LW^+_\mu+ \bar l^c_L\gamma^\mu\nu_{lL} V^+_\mu+\bar l^c_L\gamma^\mu l_L
U^{++}_\mu+H.c.\right).
\label{ccl}
\end{equation}

In order to generate the quark masses, it is necessary to introduce the
following Higgs scalars,
\begin{equation}
\eta=\left(\begin{array}{c}
\eta^0 \\ \eta^-_1 \\ \eta^+_2 \end{array}\right),\;
\rho=\left(\begin{array}{c}
\rho^+ \\ \rho^0 \\ \rho^{++}\end{array}\right),\;
\chi=\left(\begin{array}{c}
\chi^- \\ \chi^{--} \\ \chi^0\end{array}\right),
\label{scalars}
\end{equation}
transforming, under $SU(3)\otimes U(1)$, as $({\bf3},0),({\bf3},1)$
and $({\bf3},-1)$, respectively.

The lepton mass term transform as
$({\bf3}\otimes{\bf3})={\bf3}^*\oplus{\bf6}_S$, then we can introduce
a triplet, like $\eta$, but in this case one of the charged leptons
remains massless and the other two are mass degenerate, or a symmetric
antisextet $S=({\bf6}^*_S,0)$. We choose the latest one~\cite{fhpp}
in order to obtain arbitrary mass for leptons.

The charge assignment of $({\bf6}^*,0)$ is the following:
\begin{equation}
S=\left(\begin{array}{ccc}
\sigma^0_1 & h_2^+ & h_1^- \\
h_2^+ & H_1^{++} & \sigma_2^0 \\
h_1^- &\sigma_2^0 &H_2^{--}
\end{array}\right).
\label{sextet}
\end{equation}
The quark-Higgs interaction is
\begin{eqnarray}
{\cal L}_Y&=&\bar Q_{1L}(G^u_{1\alpha}U_{\alpha R}\eta+G^d_{1\alpha}
D_{\alpha R}\rho+G^jJ_{1R}\chi)\nonumber \\ \mbox{} & &
+\bar Q_{iL}(F^u_{i\alpha}U_{\alpha R}\rho^*
+F^d_{i\alpha}D_{\alpha R}\eta^*+F^j_{ik}J_{k R}\chi^*)+H.c.
\label{yuq}
\end{eqnarray}
where $\alpha=1,2,3$, $i,k=2,3$, and $U_{\alpha R}=u_{1R},u_{2R},u_{3R}$,
$D_{\alpha R}=d_{1R},d_{2R},d_{3R}$ and all fields are still
symmetry eigenstates. Explicitly from Eq.~(\ref{yuq}) one has,
\begin{eqnarray}
-{\cal L}_{QY}&=&G^u_{1\alpha}(\bar u_{1L}\eta^0+\bar
d_{1L}\eta_1^-+\bar J_{1L}\eta_2^+)U_{\alpha R}
\nonumber \\ & &\mbox{}
+G^d_{1\alpha}(\bar u_{1L}\rho^++\bar d_{1L}\rho^0+\bar
J_{1L}\rho^{++})D_{\alpha R}\nonumber \\ & &\mbox{}
+G^j(\bar u_{1L}\chi^-+\bar d_{1L}\chi^{--}+\bar J_{1L}\chi^0)J_{1R}
\nonumber \\ & &\mbox{}
+F^u_{i\alpha}(\bar J_{iL}\rho^{--}+\bar u_{iL}\rho^{0*}+\bar
d_{iL}\rho^-)u_{\alpha R}\nonumber \\ & &\mbox{}
+F^d_{iL}(\bar J_{iL}\eta_2^-+\bar u_{iL}\eta_1^++\bar
d_{iL}\eta^{0*})D_{\alpha R}\nonumber \\ & &\mbox{}
+F^j_{iL}(\bar J_{iL}\chi^{0*}+\bar u_{iL}\chi^{++}+
\bar d_{iL}\chi^+)J_{k R}+H.c.
\label{yuq2}
\end{eqnarray}
The Yukawa interaction in the leptonic sector is
\begin{equation}
{\cal L}_S=-\frac{1}{2}\sum_{ab}G_{ab}\bar\psi^c_{iaL}\psi_{jbL}S^{ij},
\label{e4}
\end{equation}
with $\psi^c=C\bar\psi^T$, being $C$ the charge conjugate matrix.
Explicitly we have
\begin{eqnarray}
2{\cal L}_S&=&-\sum_{ab}G_{ab}[\bar\nu^c_{aL}\nu_{bL}\sigma_1^0+\bar
l^c_{aL}l_{bL}H_1^{++}+ \bar l_{aR}l^c_{bL}H_2^{--}
+(\bar\nu^c_{aR}l_{bL} +\bar l^c_{aR}\nu_{bL})h_2^+\nonumber \\ & &\mbox{}
+(\bar\nu^c_{aR}l^c_{bL}+\bar l_{aR}\nu_{bL})h_1^-
+(\bar l^c_{aR}l^c_{bL}+\bar l_{aR}l_{bL})\sigma_2^0]+H.c.
\label{yul}
\end{eqnarray}
If we impose that $\langle \sigma^0_1\rangle=0$, then the neutrinos
remain massless, at least at tree level.

As we said in the last section, let us define the {\em lepto-baryon}
number, which is additively conserved, as follows
\begin{equation}
F=L+B,
\label{t}
\end{equation}
where $L$ is the total lepton number $L=\sum_iL_i,\,i=e,\mu,\tau$ and
$B$ is the baryon number.
As usual $B(l)=0$ for any lepton $l$, $L(q)=0$ for any quark $q$
\begin{equation}
F(l)=F(\nu_l)=+1,
\label{l}
\end{equation}
and
\begin{equation}
F(u_\alpha)=F(d_\alpha)=\frac{1}{3},\quad
F(J_1)=-\frac{5}{3},\;\;F(J_i)=\frac{7}{3},
\label{b}
\end{equation}
where $\alpha=1,2,3$ and $i=2,3$. From Eqs.~(\ref{l})
and (\ref{b}) we see that if
\begin{equation}
F(V^-)=F(U^{--})=+2,
\label{lb}
\end{equation}
the interactions (\ref{ccq1}), (\ref{ccq2}) and (\ref{ccl}) conserve
$F$.

In order to have $F$ also conserved in the Yukawa sector,
Eqs.~(\ref{yuq2}) and (\ref{yul}), we also
assign to the scalar fields the following values
\[
-F(\eta^+_2)=F(\rho^{++})=-F(\chi^{--})=+2,
\]
\begin{equation}
-F(H^{--}_2)=F(H^{++}_1)=F(h^+_2)=F(\sigma^0_1)=-2,
\label{lbe}
\end{equation}
and with all the other scalars fields carrying $F=0$.

Although the process $W^-V^-\rightarrow e^-e^-$ occurs in this kind
of model with the exchange of massless neutrinos, it does not imply
the $(\beta\beta)_{0\nu}$ decay since the vertex $\bar d_L\gamma^\mu
u_LV^+_\mu$ is forbidden by the $F$-conservation. This symmetry also
forbids a mixing between $W^-$ and $V^-$. Hence, we see that the $F$
symmetry must be broken in order to allow the $(\beta\beta)_{0\nu}$
to occur.

\section{The Scalar Sector}
\label{sec:scalar}
Let
\begin{eqnarray}
V(\eta,\rho,\chi,S)&=&\mu_1^2\eta^\dagger\eta+\mu^2_2\rho^\dagger\rho+
\mu^2_3\chi^\dagger\chi+\mu^2_4Tr(S^\dagger S)
+\lambda_1(\eta^\dagger\eta)^2 +\lambda_2(\rho^\dagger\rho)^2\nonumber
\\ & &\mbox{}
+\lambda_3(\chi^\dagger\chi)^2
+(\eta^\dagger\eta)\left[\lambda_4(\rho^\dagger\rho )
+\lambda_5(\chi^\dagger\chi)\right]
+\lambda_6(\rho^\dagger\rho)(\chi^\dagger\chi)\nonumber \\ & &\mbox{}
+\lambda_7\left[Tr(S^\dagger S)\right]^2
+\lambda_8 Tr(S^\dagger S S^\dagger S)+Tr(S^\dagger S)[\lambda_9
(\eta^\dagger\eta)\nonumber \\  & &\mbox{} +\lambda_{10}(\rho^\dagger\rho)
+\lambda_{11}(\chi^\dagger\chi)]
+\lambda_{12}(\rho^\dagger\eta)
(\eta^\dagger\rho)+\lambda_{13}(\chi^\dagger\eta)(\eta^\dagger\chi)
\nonumber \\  & &\mbox{}
+\lambda_{14}(\rho^\dagger\chi)(\chi^\dagger\rho)
+(f_1\varepsilon^{ijk}\eta_i\rho_j\chi_k+\frac{1}{2}f_2\rho_i\chi_j
S^{\dagger ij}+ f_3\eta_i\eta_jS^{\dagger ij}
\nonumber \\ & &\mbox{}
+\frac{1}{3!}f_4\epsilon^{ijk}\epsilon^{lmn}S_{il}S_{jm}S_{kn}+H.c.).
\label{potential}
\end{eqnarray}
This is the most general $SU(3)\otimes U(1)$ gauge invariant,
renormalizable Higgs potential for the three triplets and the sextet.
The constants $f_i,\,i=1,2,3,4$ have dimension of mass. It is
possible to show that the potential (\ref{potential}) has a local
minimum at the following vacuum expectation values (VEV) for the
scalar neutral fields~\cite{pt}
\begin{equation}
\langle\eta\rangle=\left(
\begin{array}{c}
v_\eta \\ 0\\ 0\end{array}\right),\langle\rho\rangle=\left(
\begin{array}{c}
0 \\ v_\rho \\ 0\end{array}\right),\langle\chi\rangle=\left(
\begin{array}{c}
0 \\ 0 \\ v_\chi\end{array}\right),
\label{vev}
\end{equation}
and
\begin{equation}
\langle S\rangle=\left(
\begin{array}{ccc}
0 & 0 & 0 \\
0 & 0 & v_H\\
0 & v_H & 0
\end{array}\right).
\label{vev2}
\end{equation}
Since we have chosen
$\langle\sigma^0_1\rangle=0$, the neutrinos do not gain mass at tree
level. However, we can verify the {\em naturalness} of this choice.
The situation is similar when a triplet is added to the Standard
Model~\cite{cl}. We will return to this point later.

Redefining all neutral scalars as
$\varphi=v_\varphi+\varphi_1+i\varphi_2$, except for $\sigma^0_1$,
we can analyze the scalar spectrum. For simplicity we will not
consider relative phases in the vacuum expectation values.
Requiring that the shifted potential has no linear terms in any of
the $\varphi_{1,2}$ components of all neutral scalars we obtain in
the tree approximation the constraint equations:
\begin{equation}
\begin{array}{r}
\mu^2_1+2\lambda_1v^2_\eta+\lambda_4v^2_\rho+\lambda_5v^2_\chi+
2\lambda_9v^2_H+f_1v^{-1}_\eta v_\rho v_\chi=0,\\
\mu^2_2+2\lambda_2v^2_\rho+\lambda_4v^2_\eta+\lambda_6v^2_\chi+
2\lambda_{10}v^2_H+f_1v_\eta v^{-1}_\rho v_\chi+f_2v_\chi v_Hv_\rho^{-1}=0,\\
\mu^2_3+2\lambda_3v^2_\chi+\lambda_5v^2_\eta+\lambda_6v^2_\rho+
2\lambda_{11}v^2_H+f_1v_\eta v_\rho v^{-1}_\chi+f_2v_\rho v_Hv_\chi^{-1}=0,\\
\mu^2_4+4\lambda_7v^2_H+2\lambda_8v^2_H+
\lambda_9v^2_\eta+\lambda_{10}v^2_\rho+\lambda_{11}v^2_\chi+
f_2v_\rho v_\chi v^{-1}_H=0,\\
f_3v^2_\eta-f_4v^2_H=0,\\
Im\,f_i=0,\quad i=1,2,3,4,
\end{array}
\label{ce}
\end{equation}
and the mass matrix in the
$\eta^-_1,\eta^-_2,\rho^-,\chi^-,h^-_1,h^-_2$ basis is
\begin{equation}
v^2_\chi\left(
\begin{array}{cccccc}
A_1  \, & A_2\,    & -F_3c\;   & 0 \;     & 0 \,     & -F_3a\, \\
A_2\,   & A_3\,    &  0 \,    & 0 \,     & -F_2\,   & 0\,     \\
-F_3c\, & 0\,      & B_1\,    & B_2\,    & -F_3a\,   & 0\,      \\
0 \,    & 0\,      & B_2\,    & B_3\,    & 0\,      & -F_2b\,   \\
0 \,    & -F_2\;   & -F_3a\;  & 0\,      & C_1\,    & C_2 \, \\
-F_3a\, & 0\,      & 0     \, & -F_2b\,  & C_2\,    & C_1\,
\end{array}\right)
\label{mm}
\end{equation}
Where
\begin{equation}
A_1=F_1ba^{-1}-\lambda_{12}b^2,\quad A_2=F_1-\lambda_{12}ab,
\quad A_3=(F_1a+F_2c)b^{-1}-\lambda_{12}a^2,
\label{def1}
\end{equation}
\begin{equation}
B_1=F_1ba^{-1}-\lambda_{13},\quad B_2=F_1b-\lambda_{13}a,
\quad B_3=(F_1a+F_2c)b-\lambda_{13}a^2,
\label{def2}
\end{equation}
\begin{equation}
C_1=F_2bc^{-1},\quad C_2=-F_3a^2c^{-1},
\label{def3}
\end{equation}
and we have defined the dimensionless constants $F_i=f_1/v_\chi$,
$a=v_\eta/v_\chi$, $b=v_\rho/v_\chi$ and $c=v_H/v_\chi$.
The mass matrix in Eq.~(\ref{mm}) has just two Goldstone bosons and
implies a mixing among all charged scalars, hence the physical
charged scalars are linear combinations of
$\eta^-_i,h^-_i(i=1,2)$, $\rho^-$  and $\chi^-$ which have no well
defined value of $F$. As quarks
$u,d$ interact according Eq.~(\ref{yuq2}) with $\eta^-_1,\rho^-$
and the last fields are linear
combinations of mass eigenstates we see that the diagram in
Fig.~1 is possible even if the neutrino were massless.

\section{The $(\beta\beta)_{0\nu}$ Decay}
\label{sec:decay}
The $F$ symmetry is softly broken by the $f_{3,4}$ terms
in the scalar potential (\ref{potential}). As we said in the last
section, the physical singly charged scalar, in the present case, are
not eigenstates of $F$. Then, if $\Phi^-_1$ in Fig.~1 is one of the
scalar mass eigenstates, in general
\begin{equation}
\phi^-_i=\sum_{ij}a_{ij}\Phi^-_j,
\label{phi}
\end{equation}
with $\phi^-_i=\eta_1^-,\eta^-_2,\rho^-,\chi^-,h^-_1,h^-_2$ and
$a_{ij}$ are mixing parameters.

We can estimate a lower bound on the mass
of $\phi^-_1$, by assuming that its contribution to
$(\beta\beta)_{0\nu}$ is less than the amplitude due to massive
Majorana neutrinos and vector bosons $W^-$ exchange. The latter
amplitude is characterized by a strength which is proportional to
\begin{equation}
\frac{g^4m^{eff}_\nu}{m^4_W<p^2>},
\label{e10}
\end{equation}
where $m^{eff}_\nu=\vert\sum_j U^2_{lj}m_j\vert$ is the ``effective
neutrino mass''~\cite{d}. The experimental limit on $(\beta\beta)_{0\nu}$
decay rate imply that $m^{eff}_\nu\,<
M_\nu=(1-2)\,\mbox{eV}$~\cite{bk}; $<\!\!p^2\!\!>$ is an average
square 4-momentum carried by the virtual neutrino, its value is
usually $(10 MeV)^2$~\cite{sv}. On the other hand, the amplitude of
the process in Fig.~1 is proportional to
\begin{equation}
\frac{(a_{11}a_{21})^2G_{ud}^4G^4_{ee}}{ m^4_\phi<p^2>^{\frac{1}{2}}},
\label{e11}
\end{equation}
Then, assuming that Eq.~(\ref{e11}) is less than Eq.~(\ref{e10}) when
$m^{eff}_\nu=M_\nu$ we have,
\begin{equation}
m^4_{\phi_1}>\frac{(a_{11}a_{21})^2G^4_{ud}G^4_{ee}\sqrt2<p^2>^{\frac{1}{2}}}
{32G_F^2M_\nu}\simeq (6,9\,\mbox{ TeV})^4(a_{11}a_{21})^2G^4_{ud}G^4_{ee}.
\label{e12}
\end{equation}
The factors in Eq.~(\ref{e11}) arise as follows. In Eq.~(\ref{yuq2})
the fields are symmetry eigenstates. It is possible to redefine the
quark fields as
\begin{equation}
q'_L=V^Q_Lq_L,\quad q'_R=V^Q_Rq_R
\label{masse}
\end{equation}
with $V^Q_{L,R}$ being unitary matrices in the flavor space and the
primed fields denote mass eigenstates for the respective
charge-$Q$ sector. In Eq.~(\ref{yuq2}) the interactions are as
$G_{ud}\bar d_Lu_R\eta_1^-$ with
$G_{ud}=(V_L^{(-\frac{1}{3})}G^uV_R^{(\frac{2}{3})})_{ud}$ and $d,u$
are mass eigenstates. The coefficients $G_{ee}$ appear in
Eq.~(\ref{yul}). As these mixing parameters in Eq.~(\ref{e12}) can be
very small it does not imply a strong lower bound on the scalar fields.

There are not contributions to $(\beta\beta)_{0\nu}$ from trilinear
Higgs interactions like $\eta^-_1h^-_2H^{++}$. In models in which
these contributions exist, they are negligible~\cite{sv} unless a
neighboring mass scale $(\sim 10^4\, GeV)$ exist~\cite{ep}.
\section{Conclusions}
\label{sec:con}
As promised, we now consider the question of the neutrino masses.
First at all, notice that if we forbid, by assuming a discrete
symmetry~\cite{fhpp} the trilinear terms in $f_{3,4}$ the mixing arisen
from Eq.~(\ref{mm}) is among $\eta^-_1,\rho^-,h^-_1$ and separately
among $\eta^-_2,\chi^-,h^-_2$ and the $F$ is conserved and
$(\beta\beta)_{0\nu}$ is forbiden. Then, it is important the  fact that
the $F$ symmetry is softly broken by the trilinear terms $f_{3,4}$
in the scalar potential (\ref{potential}).

As we have made $\langle \sigma^0_1\rangle=0$, we can think that the
neutrino masses vanish at tree level but that they are finite and
calculable, in the sense of Sec.~\ref{sec:intro}. Let us consider
this issue more in detail.

Besides (\ref{yul}) there is the additional Yukawa coupling between
leptons and the triplet scalar $\eta$,
\begin{equation}
{\cal
L}_{l\eta}=\sum_{ab}f_{ab}\bar\psi^c_{aiL}\psi_{bjL}\epsilon^{ijk}\eta_k+H.c.,
\label{yul2}
\end{equation}
$a,b$ denote family indices, $i,j$ denote $SU(3)$ indices and
$\epsilon^{ijk}$ is the totally antisymmetric symbol. The Yukawa
couplings $f_{ab}$ must be antisymmetric, $f_{ab}=-f_{ba}$, due to
Fermi statistics and the antisymmetric property of the charge
conjugation matrix, $C$. Then, Eq.~(\ref{yul2}) connects leptons of
different families. Typical terms of Eq.~(\ref{yul2}) read
\begin{equation}
(\bar\nu^c_{eL}\mu^-_L-\bar e^c_L\nu_{\mu L})\eta^+_2,\;
(\bar\nu^c_{eL}\mu^+_L-\bar e_L\nu_{\mu L})\eta^-_1.
\label{yul3}
\end{equation}
Next, due to Eqs.~(\ref{yul}) and (\ref{yul3}), the neutrinos gain
finite masses through loop diagrams as in Figs.~2(a) and
2(b).

Now, joining the neutrino lines in Figs.~2 by
$\sigma^0_1$, we obtain the divergent contribution to
$\langle\sigma^0_1\rangle$ appearing in
Fig.~3(a,b). This implies a counterterm and makes it impossible to
maintain $\langle\sigma^0_1\rangle=0$, at least in a {\em natural}
way. Hence, neutrinos gain an arbitrary small mass since we can always
to assume that $\langle\sigma^0_1\rangle\simeq0$.

If $\langle\sigma^0_1\rangle\not=0$ there is
a mixing in the charged vector sector $W^+$-$V^+$,
\begin{equation}
W^\pm=\alpha X^\pm_1+\beta X^\pm_2,\quad V^\pm=-\beta X^\pm_1+\alpha
X^\pm_2, \quad \alpha^2+\beta^2=1,
\label{x}
\end{equation}
where $X^\pm_{1,2}$ are mass eigenstates. Hence the
$(\beta\beta)_{0\nu}$ proceeds also as in Fig.~4, whithout a direct
dependence on the neutrino mass, but this contribution is suppressed
by the large mass of the vector boson $X_2^-$ or by the mixing
parameters since $\beta\simeq0$. If
$\langle\sigma^0_1\rangle\not=0$, and assuming discrete symmetries to
forbid the explicit violations in Eq.~(\ref{potential}), we have
spontaneously breakdown of the $F$ symmetry, since $\sigma^0_1$
carries $F=-2$, see Eq.~(\ref{lbe}), implying a like-Majoron Goldstone
boson since the $\sigma^0_1$ belongs to a triplet under $SU(2)$
together with $h^-_2$ and $H^{--}_1$. The phenomenology of this Goldstone
boson can be similar to that of the Majoron~\cite{gr} and it deserve
a more detail study. As $ \langle\sigma^0_1\rangle\not=0$ we expect a
deviation from the $\rho=1$ value ($\rho=\cos^2\theta_WM^2_Z/M^2_W$),
however as $ \langle\sigma^0_1\rangle$ is arbitrary small its
contributions can be made negligible.

Summarizing: we see that $(\beta\beta)_{0\nu}$ proceeds in this
model also as a Higgs bosons effect, with almost massless neutrinos.
Recall that if we have forbidden all trilinears in
Eq.~(\ref{potential}) except those with $f_1,f_2$, neutrinos could
remain massless but the mixing in the
charged scalar sector is only among $\eta^-_1,\rho^-,h^-_1$ and
$\eta^-_2,\chi^-,h^-_2$ separately. Hence the $(\beta\beta)_{0\nu}$
cannot occur as was shown in Ref.~\cite{pp}.

In this sort of model, there must be exotic hadrons formed by
combinations such as $qqJ_2$ with $B=3$, being $q$ any of the known
quarks.

\acknowledgements

We would like to thank the Con\-se\-lho Na\-cio\-nal de
De\-sen\-vol\-vi\-men\-to Cien\-t\'\i \-fi\-co
e Tec\-no\-l\'o\-gi\-co (CNPq) for full (M.D.T.) and partial (V.P.)
financial support. One of us (V.P.) is indebted to G.B. Gelmini for
enlightening discussions at the beginning of this work.

{\bf \center Figure Captions \\}

\noindent {\bf Fig. 1} Scalar contribution to the $(\beta\beta)_{0\nu}$,
$G_{ud,ee}$ are Yukawa couplings and $a_{11,21}$ mixing parameters in
the scalar sector.

\noindent {\bf Fig. 2} Finite contribution to the Majorana neutrino
mass due to the Yukawa couplings in Eqs.~(\ref{yul}) and (\ref{yul3})
and the trilinear terms $f_3$ and $f_4$ in the scalar potential
(\ref{potential}).

\noindent {\bf Fig. 3} Joining the neutrino lines by $\sigma_1^0$ in
Fig.~2 we obtain a divergent contribution to $\langle
\sigma^0_1\rangle$.

\noindent {\bf Fig. 4} Contribution to the $(\beta\beta)_{0\nu}$ due
the charged vector bosons exchange if
$\langle\sigma^0_1\rangle\not=0$. $X^+_1$ is a
linear combination of $W^+$ and $V^+$. See Eq.~(\ref{x}).
\end{document}